\documentclass[10pt]{amsart}

\usepackage[margin=1in]{geometry}
\usepackage{graphicx}
\usepackage{amsmath}
\usepackage{amssymb}
\usepackage{amscd}
\usepackage{amsthm}

\newtheorem{remark}{Remark}
\newtheorem{definition}{Definition}
\newtheorem{proposition}{Proposition}

\newcommand{\Hom}{\operatorname{Hom}}
\newcommand{\Ext}{\operatorname{Ext}}

\title{Homological Coordinatization}

\author{Andrew Tausz}
\thanks{The first author was supported by AFOSRG grant FA9550-09-0-1-0531.}
\address{Stanford University, Stanford, CA, 94305}
\email{atausz@stanford.edu}

\author{Gunnar Carlsson}
\thanks{The second author was supported by AFOSRG grant FA9550-09-0-1-0531, ONR grant N00014-08-1-0931 and NSF grant DMS 0905823.}
\address{Stanford University, Stanford, CA, 94305}
\email{gunnar@math.stanford.edu}

\date{\today}

\begin{document}

\maketitle

\begin{abstract}

In this paper, we review a method for computing and parameterizing the set of homotopy classes of chain maps between two chain complexes. This is then applied to finding topologically meaningful maps between simplicial complexes, which in the context of topological data analysis, can be viewed as an extension of conventional unsupervised learning methods to simplicial complexes.

\end{abstract}

\section{Introduction}

One goal of topological data analysis is to adapt algebraic topological methods to the context of point cloud data (i.e. finite metric spaces). The generalization of  homology to this setting is called {\em persistent homology} \cite{Carlsson_04}, \cite{Carlsson_09}. Persistent homology has been successful at providing insight into nonlinear datasets, that would not be accessible with more classical methods. Although maps between spaces play a fundamental role in algebraic topology, most of the developments in topological data analysis have focused on the spaces and datasets themselves. In this paper, we propose a method for studying these maps.

Suppose that we compute the homology of a space $X$. By comparing the homology of $X$ with the known homology of other spaces, it can suggest that $X$ is homeomorphic to a previously understood space, or that there should be maps exhibiting prescribed homological behavior to model spaces with known homology.   One thinks of this process as a kind of non-linear coordinatization. Standard methods for introducing useful coordinates on a point cloud include Principal Component Analysis and Multidimensional Scaling.  Both of these methods often work well when the structure of the space is essentially Euclidean.  However, when the space carries noncontractible topology, as in the case of a circle, these methods do not provide a method for mapping the data set to a nonlinear model.  We describe some examples when the ability to construct maps to nonlinear targets would be useful.

{{\bf Circular coordinatization:}  In situations where one finds that the persistent homology of the data set is that of a circle, it is natural to attempt to find a map to the circle.  In this special case, there is a natural methodology using the persistent cohomology generator in dimension 1 which allows one to construct the map.  The procedure is described in detail in \cite{de_Silva_Johansson}.   }

{{\bf Natural image example:} In \cite{mumford}, homological calculations where carried out to confirm that a space of frequently occurring motifs within $3 \times 3$ image patches in natural images had the homology of a Klein bottle. Once one is given the homology, it is then of a great deal of interest to attempt to construct an actual parametrization by a Klein bottle.  This was done by hand in \cite{mumford}, but one would like to automate the procedure.}

{{\bf Gene expression data:}  Gene expression microarrays provide a powerful tool for obtaining information about many biological phenomena, including cancer.  They produce high dimensional data, where the coordinates  consist of probes representing particular genes.   It is very well known that the results of such studies are highly dependent on platforms, procedures within the laboratories performing the studies, as well as many other factors.  All this can distort the geometry of the data set, but one can hope that certain topological features would still be preserved, which might permit one to map one data set to another in a nonlinear way, preserving the relevant features.  Often \cite{monica}, the geometry of these data sets are represented by shapes like the one pictured below.

\hspace{4.5cm}\includegraphics[height=5cm]{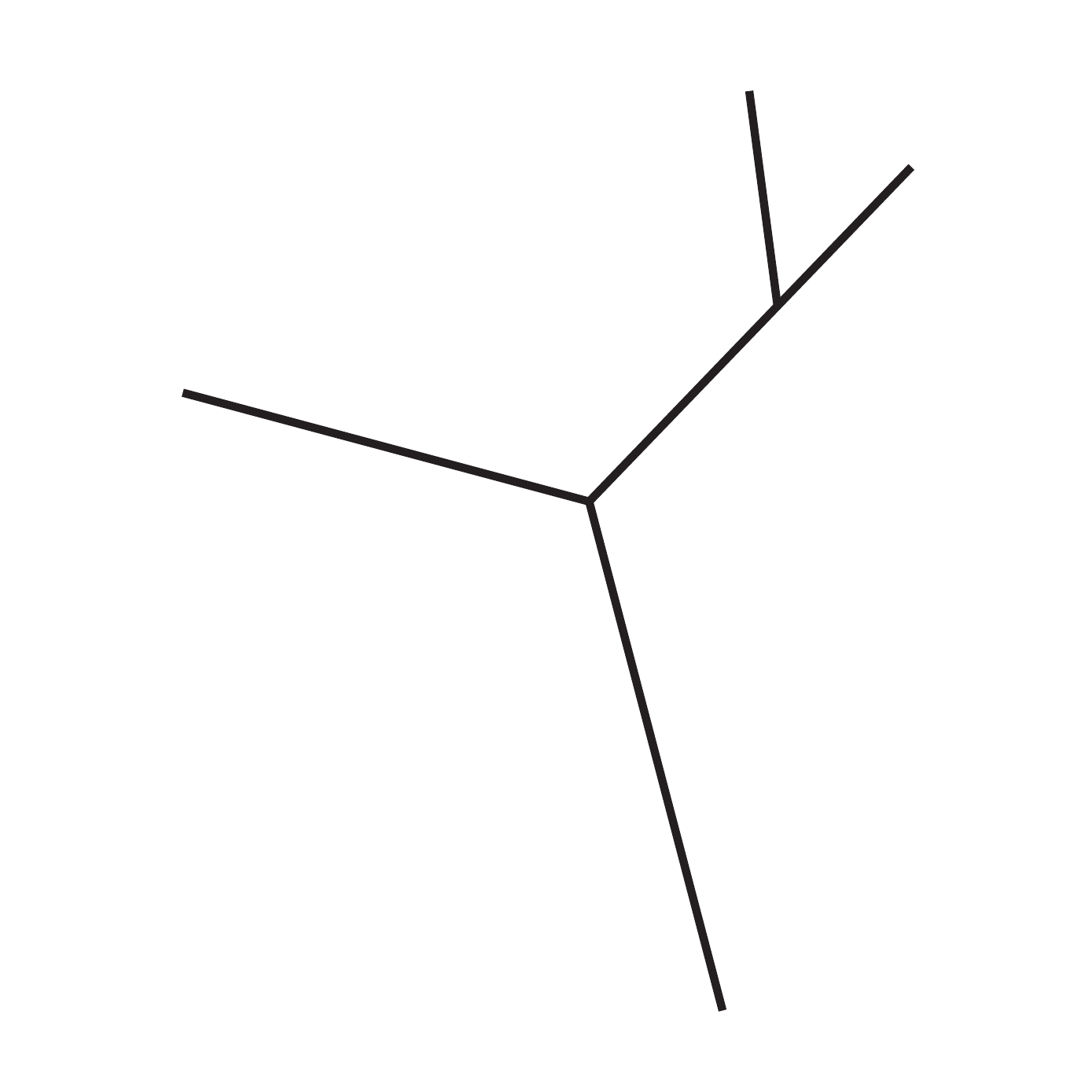}

In this case, since the tree is contractible, the direct use of homology will not be useful, since homology vanishes on contractible spaces.  However, {\em relative homology} of the pair $(X, \partial X)$, where $\partial X$ denotes a suitably defined boundary of the space, does capture the existence of branches or flares in the geometric object. There are reasonable ways of defining the boundary of the metric space, and the points in the boundary have significance in that they tend to consist of most representative phenomena of a particular subclass of samples.  For example, type I and type II diabetes can be distinguished in this way, as can various molecular subtypes of cancer.  In this case, one could then fix the map on the boundary, and study the relative mapping problem in which one enforces the constraint that the boundary is carried into a small neighborhood of the boundary.  This kind of mapping is of a great deal of importance, since the problem of reconciling different data sets constructed on the same kinds of tumors or other disorders is a major problem in this area.  }

From the perspective of topological data analysis, we desire a map between two simplicial complexes $X$ and $Y$ to satisfy the following:
\begin{itemize}
\item Such a map must be functorial: It must induce homomorphisms on the homology of $X$ and $Y$.
\item Ideally, the map would be {\em simplicial}. This means that the image of a simplex should be a chain in $Y$ containing either 0 or 1 simplices. Simplicial maps are particularly very well behaved: they are determined by their values on vertices and can be fully extended by linear interpolation. Additionally, they have a nice geometric interpretation. In general, we obtain maps between {\em chains} in $X$ and $Y$. Thus, the typical situation is that we have $f(\sigma) = \sum_{\tau_j \in Y, |\tau_j| = |\sigma|} a_j \tau_j$. 
\item Even though a map might be a chain map (hence inducing homomorphisms, or even isomorphisms in homology), such a map might be unsatisfactory in the eyes of a practitioner. We want these mappings to reveal some sort of information about the structure of one complex in terms of the other. A common situation might be that $X$ is created from a large and high-dimensional dataset, and $Y$ is a simple model with the same homology. In this case, a map $X \rightarrow Y$ can be thought of as performing a kind of topological dimensionality reduction. This leads to the process of geometric regularization through optimization.
\end{itemize}

Unfortunately, simplicial maps do not always exist - an example is the case of mapping from a triangle to a square. For this reason, we wish to find maps that are as close to simplicial as possible. Our investigation to this problem proceeds as follows:

\begin{itemize}
\item The first step is to compute a compute a parameterization of the homotopy classes of chain maps between two finite simplicial complexes. This is accomplished quite easily using a simple trick from homological algebra. (See sections \ref{section_basics} and \ref{section_maps})
\item We wish to develop optimization problems that yield maps that are as close to simplicial as possible. Additionally, we would like these problems to satisfy other properties. For example, when they exist, simplicial maps should be contained in the set of optima, and to preserve tractability the problem should be convex. We discuss these criteria in section \ref{section_optimization} and formulate two different optimization problems over the parameterization. The first one is convex and can be formulated as a linear program, while the second is non-convex but has the property that its minima are precisely the set of simplicial maps when they exist. 
\item In section \ref{section_applications}, we discuss three applications to various scenarios in topological data analysis: manifold-valued coordinates, density maximization, and mappings of contractible spaces. In the third example, we show how to compute mappings between trees by considering their relative homology with respect to a boundary defined by a filter function.
\end{itemize}

The two main existing investigations into the area of computing maps are \cite{de_Silva_Johansson} and \cite{Ding}. In the paper \cite{de_Silva_Johansson}, de Silva and Vejdemo-Johansson present a method based on persistent cohomology for computing circular-valued coordinates on statistical data sets. The fundamental idea behind their work is to use the Brown representation of cohomology. Since we know that $S^1$ is the Eilenberg-MacLane space for the group $\mathbb{Z}$, we have that
$$H^1(X; \mathbb{Z}) \cong [X, K(\mathbb{Z}, 1)] = [X, S^1]$$
We can compute a map from a space $X$ to the circle $S^1$ by choosing a representative from a cohomology class $H^1(X; \mathbb{Z})$. In practice, they also perform an optimization step where they select the smoothest cocycle. Although this works very well for the case of a circle, it is not practically generalizable to spaces other than $S^1$. For example $K(\mathbb{Z}, 2)$ is the infinite-dimensional complex projective space $\mathbb{CP}^{\infty}$, and $K(\mathbb{Z}/2\mathbb{Z}, 1)$ is $\mathbb{RP}^{\infty}$.

In the PhD thesis of Yi Ding, \cite{Ding}, the mapping problem is investigated from a combinatorial perspective. The hom-complex is used, as it is here, to obtain a parameterization of the space of chain maps, but combinatorial optimization techniques are used to select maps that satisfy certain criteria. Our work can be seen as an extension of Ding's work to the continuous case.

We close the introduction with a remark on how we can think about the simplicial mapping problem as a version of higher order clustering. Clustering can be thought of as computing a map from a dataset $X$ to the discrete space $Y = \{1, ..., k\}$. In the figure below, a cluster assignment is a mapping from each point in the set on the left to the set $\{1, 2, 3\}$.

\hspace{2.5cm}\includegraphics[height=5cm]{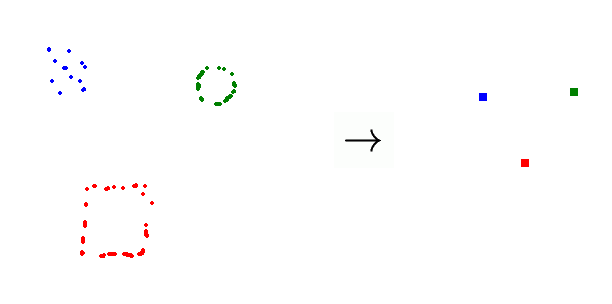}

Suppose that we construct a filtered simplicial complex from $X$ with some maximum filtration parameter $r_{\max}$. Examples of such constructions include the Vietoris-Rips complex, the Cech complex, and others. In this case, a clustering assignment is a mapping from the filtered complex to $\{1, ..., k\}$ that is constant on the path components of the complex. It is easy to verify that this is equivalent to the chain map property on dimension 0: If $e$ is an edge between $u$ and $v$ in the same path component, then $f(u) - f(v) = f(\partial e) = \partial f(e) = 0$. Thus, a higher dimensional analogue of clustering is the computation of some homotopy representative of a class of chain maps for $n$-simplices, with $n > 0$.

\section{Definitions and Basics}
\label{section_basics}

In this section we review some basic definitions for completeness. This material can be found in a standard text on algebraic topology such as \cite{HATCHER} or \cite{Munkres}. The material on the hom-complexes can be found in \cite{Maclane}.

\begin{definition}
A chain map between two chain complexes $(A_n, d_n)$ and $(B_n, d'_n)$ is a family of homomorphisms $f_n: A_n \rightarrow B_n$ such that for each $n$ we have that $f_{n-1} d_n = d'_n f_n$. 
\end{definition}

\begin{definition}
Given two chain maps $f$ and $g$ between the chain complexes $(A_n, d_n)$ and $(B_n, d'_n)$, a chain homotopy is a family of homomorphisms $s_n: A_n \rightarrow B_{n+1}$ such that for each $n$ we have that 
\begin{equation}
d'_{n+1} s_n + s_{n-1} d_n = f_n - g_n
\end{equation}
\end{definition}

The key fact is that if the chain maps $f$ and $g$ are chain-homotopic, then they induce the same homomorphism on homology.

\begin{definition}
Given two chain complexes $(A_*, d_*)$ and $(B_*, d'_*)$, we define a new complex as follows. Let
\begin{equation}
\Hom_n(A_{*}, B_{*}) = \bigoplus_{p = -\infty}^{\infty} \Hom(A_p, B_{p+n})
\end{equation}
\end{definition}

An element $f$ of $\Hom_n(A_{*}, B_{*})$ is a family of homomorphisms $f_p:A_p \rightarrow B_{p+n}$ for $p \in \mathbb{Z}$. Note that even in the case where the chain complexes $(A_*, d_*)$ and $(B_*, d'_*)$ have non-negative support, the chain complex $\Hom_n(A_{*}, B_{*})$ will have non-trivial negative terms in general.

Now that we have defined the terms in the complex, we need to define connecting homomorphisms
$$d^H_n: \Hom_n(A_{*}, B_{*}) \rightarrow \Hom_{n-1}(A_{*}, B_{*})$$
If $f \in \Hom_n(A_{*}, B_{*})$, then $d^H_n(f)$ will itself be a family of homomorphisms. Let us denote by $f_p$, the component mappings on the individual modules $A_p$. To define $d^H_n(f)$, it is sufficient to specify its action on elements of $A_p$ for all $p$. Let $a \in A_p$, be an arbitrary element. We define
$$d^H_n(f)(a) = d'_{p+n}(f_p(a)) + (-1)^{n+1} f_{p-1} (d_p(a))$$

It is easy to see that $d^H$ satisfies $d^H_n d^H_{n+1} = 0$. We summarize the properties of the hom-complex in the proposition below. Each claim can be proved by a simple computation.

\begin{proposition}
The hom-complex satisfies the following homological properties:
\begin{enumerate}
\item  $f \in \Hom_0(A_{*}, B_{*})$ is a cycle ($d^H(f) = 0$) if and only if $f$ is a chain map.
\item $f \in \Hom_0(A_{*}, B_{*})$ is the boundary of some element $s \in \Hom_1(A_{*}, B_{*})$ if and only if $f$ is chain homotopic to the zero map via $s$.
\item The zeroth homology group $H_0(\Hom_*(A_{*}, B_{*}))$ consists of homotopy classes of chain maps.
\end{enumerate}
\end{proposition}

Thus we have a nice characterization of the homotopy classes of chain maps - all we need to do is compute the homology of the hom-complex. Note that in this paper, if we use the term homotopy we actually mean the term ``chain homotopy''. In other words, we only deal with the algebraic notion and not the topological one. Furthermore, by a map between chain complexes, we really mean a chain map.

\begin{remark}
In our investigation, we will restrict ourselves to field coefficients for two computational reasons. Firstly, the homology computations will be performed in a persistent setting for which one must work over a field (see \cite{Carlsson_04}). Secondly, working over a field of characteristic 0 greatly enhances the efficiency of the later optimization steps. As a bonus for working with field coefficients, we get a more manageable representation of the individual $\Hom$ terms, since for vector spaces $V$ and $W$ we have that $\Hom(V, W) \cong V^{\wedge} \otimes W$. We denote the vector space dual of $V$ with $V^{\wedge}$. In this case, the $n$-th term in the hom-complex is
\begin{equation}
\Hom_n(A_{*}, B_{*}) = \bigoplus_{p = -\infty}^{\infty} (A_p)^{\wedge} \otimes B_{p+n}
\end{equation}
This is very similar to another standard construction in homology theory - the tensor product of two chain complexes. Thus an element $f \in \Hom_n(A_{*}, B_{*})$ may be written as $f = \sum c_{ij }a_i^{*} \otimes b_j$. From the computational point of view, this representation is particularly useful since most of the coefficients $c_{ij}$ will be zero and do not need to be stored.
\end{remark}
We now close this section with a theorem which gives a useful characterization of the homology of the hom-complex in special cases. A proof may be found in chapter III of \cite{Maclane}. 

\begin{proposition}
\label{hclassification}
Let $(A_n, d_n)$ and $(B_n, d'_n)$ be chain complexes of $R$-modules. Then there exists the following exact sequence for each $n$

\begin{equation}
0 \rightarrow \bigoplus_{-\infty}^{\infty} \Ext_R^1(H_k(A), H_{k+n+1}(B)) \rightarrow H_n(\Hom_*(A, B)) \rightarrow \bigoplus_{-\infty}^{\infty} \Hom(H_k(A), H_{k+n}(B)) \rightarrow 0
\end{equation}

In the special case when we are dealing with vector spaces over a field $\mathbb{F}$, then the Ext term vanishes and we have an explicit expression for the Hom term. So we get that
\begin{equation}
H_n(\Hom_*(A_{*}, B_{*})) \cong \bigoplus_{-\infty}^{\infty} \Hom(H_k(A), H_{k+n}(B)) \cong \bigoplus_{-\infty}^{\infty} H^k(A) \otimes H_{k+n}(B)
\end{equation}

\end{proposition}

\section{Finding Mappings between Simplicial Complexes}
\label{section_maps}

In this section we apply the algebraic techniques of the previous section to computing a parameterization of the homotopy classes of maps between two simplicial complexes.

Suppose we have simplicial complexes, $X$ and $Y$ with $X = \{\sigma_i\}$, $Y = \{\tau_j\}$. So $\{\sigma_i\}$ and $\{\tau_j\}$ are bases for the vector spaces (over a field, $\mathbb{F}$) $C_*(X)$ and $C_*(Y)$. We also denote the dual basis of $C_*(X)$ with $\{\sigma^*_i\}$. Then, the vector space $\Hom(C_*(X), C_*(Y))$ has basis $\{\sigma_i^* \otimes \tau_j\}$.

Based on our previous discussion about hom-complexes, we know that the set of homotopy classes of maps between $X$ and $Y$ is given by the $0$-th homology $H_0(\Hom_*(C_*(X), C_*(Y)))$ with
\begin{equation}
\Hom_n(C_*(X), C_*(Y)) = \bigoplus_{p = -\infty}^{\infty} C^{p}(X) \otimes C_{p+n}(Y)
\end{equation}

In practice, we are not interested in the rank or Betti numbers of the homology group, but rather we wish to find representatives for the homotopy classes. As described in \cite{Munkres}, homology can be computed via the Smith normal form in the case of coefficients in a PID, or with Gaussian elimination in the case of coefficients in a field. 

The result of the homology computation is a set of representative cycles (equivalently chain maps), which we denote $\{f_m\}$. We can also easily obtain the set of chain homotopies to zero by computing the columns (the image) of the matrix representation of $d^H_1$. Let us denote these columns by $\{h_n\}$. 

The general parameterization of the affine space of homotopy classes is
$$[X, Y] = \left\{ \sum_m b_m f_m + \sum_n c_n h_n | b_m \in \mathbb{F}, c_n \in \mathbb{F} \right\}$$
Note that the above expression uses the notation $[X, Y]$ for the \emph{chain-homotopy classes of chain maps} between $X$ and $Y$, and not the homotopy classes of all continuous maps. Let us denote the $m$-th chain map by $f_m = \sum_{ij} f^m_{ij} \sigma_i^* \otimes \tau_j$, and the $n$-th chain homotopy by $h_n = \sum_{ij} h^n_{ij} \sigma_i^* \otimes \tau_j$.

\subsection{Example}

Let $X$ be a triangle with vertices [0], [1], [2], and edges [0, 1], [0, 2], [1, 2]. Let $Y = X$. Note that by the homotopy classification theorem of the previous section (Proposition \ref{hclassification}), we have that
$$H_0(\Hom_*(C_{*}(X), C_{*}(Y))) \cong \bigoplus_{-\infty}^{\infty} H^k(X) \otimes H_{k}(Y) \cong \mathbb{F} \oplus \mathbb{F}$$
Thus we have two generating cycles. The two computed representatives of $H_0(\Hom_n(C_{*}(X), C_{*}(Y)))$ are:
\begin{eqnarray*}
f_0 & = & [0]\rightarrow[0] + [1]\rightarrow[0] + [2]\rightarrow[0]\\
f_1 & = & -[1,2]\rightarrow[0,2] + [1,2]\rightarrow[0,1] + [1,2]\rightarrow[1,2]
\end{eqnarray*}
The set of homotopies is given by the image of the 1-dimensional boundary matrix $d^H_1$. The first few (out of a total of 9) are:
\begin{eqnarray*}
h_0 & = & [1,2]\rightarrow[0,1] + [0,2]\rightarrow[0,1] + [2]\rightarrow[1] - [2]\rightarrow[0]\\
h_1 & = & [1,2]\rightarrow[0,2] + [0,2]\rightarrow[0,2] + [2]\rightarrow[2] - [2]\rightarrow[0]\\
h_2 & = & [2]\rightarrow[2] - [2]\rightarrow[1] + [1,2]\rightarrow[1,2] + [0,2]\rightarrow[1,2]\\
\ldots
\end{eqnarray*}

We can see that the first generating cycle induces an isomorphism $H_0(X) \rightarrow H_0(Y)$, and induces the zero map on dimension-1 homology. Similarly, the second generating cycle induces an isomorphism $H_1(X) \rightarrow H_1(Y)$ and induces the zero map on dimension-0 homology. A generator for $H_0(X)$ (equivalently $H_0(Y)$) is $[0]$, and a generator for $H_1(X)$ (equivalently $H_1(Y)$) is $[0,1] - [0,2] + [1,2]$. A quick computation also reveals that adding a chain homotopy does not change the induced action of the generating cycles.

This example also gives us some hints about the selection of the coefficients $\{b_m\}$ for the chain maps. From the previous paragraphs, we have chain maps (which are homology cycles), $f_0$ which induces an isomorphism on dimension 0 homology, and $f_1$ which induces an isomorphism on dimension 1 homology. If we take their sum, $f = f_0 + f_1$ (by setting $b_0 = b_1 = 1$), then it is easy to see that $f$ induces an isomorphism on both homology groups. In practice this map is the one to start with, since it preserves the homological structure across all dimensions. Thus in this case we would use the parameterization
$$\left\{ f + \sum_n c_n h_n | c_n \in \mathbb{F} \right\}$$

\section{Searching the Parameterization}
\label{section_optimization}

Given two simplicial complexes $X$ and $Y$, we now know how to compute a parameterization for the homotopy classes of chain maps between them. However, in a statistical application setting we are interested in selecting only one geometrically meaningful map from this set. Some reasonable criteria for such a map are:

\begin{itemize}
\item Image cardinality (simplicialness): In general, the image of a simplex $\sigma$ under a map $g$ will be some linear combination $g(\sigma) = \sum_j d_j \tau_j$. Ideally, we would want to have the number of non-zero coefficients $c_j$ to be as few as possible.
\item Preimage cardinality: Likewise, we would prefer if the number of non-zero coefficients in the expression $g^*(\tau) = \sum_i c_i \sigma_i$ be as few as possible.
\item Locality: We would like the image of a simplex to be localized in the codomain. This means that the non-zero terms in the expression $g(\sigma) = \sum_j d_j \tau_j$ should not be spread apart.
\item Unbiasedness: There should be no a-priori reason to prefer one map over another if they achieve the same optima. For example, in the case of a triangle mapping to a triangle, there is no geometric way of distinguishing the identity map from one of the two rotations of the vertices (both are simplicial and optimally localized).
\item Convexity: The optimization problem should be convex.
\end{itemize}

Unfortunately, the above criteria are mutually incompatible. To see this, it suffices to consider the case where $X$ and $Y$ are both triangles. The optimally sparse and localized chain maps include the three rotations of the vertices. However, the unbiased property says that each of these three maps should be elements of the set of optima of the optimization problem. If we require the problem to be convex, then it turns out that the set of optima must also be convex and in particular connected. Thus if one takes a non-trivial convex combination (say with coefficients (1/3, 1/3, 1/3)), that will also be an optima but it will violate the condition of sparsity. One remedy for this is to require that the set of optima is the convex hull of the unbiased set of points. Alternatively, one can discard unbiasedness and require that only one sparse point be returned.

\subsection{The Combinatorial Approach is Hard: Theory}

The purpose of this section is to provide some orientation regarding the complexity of finding nice maps. This assumes that we are taking a combinatorial approach as discussed in \cite{Ding}.

Our ultimate goal would be to construct a map that is simplicial in both directions. This means that the image and preimage of each simplex contains zero or one simplices. Let us call such a map \emph{bisimplicial}. In other words, such a map would minimize the quantity
\begin{equation}
\label{bisimplicial}
\max_{\sigma \in X} ||f(\sigma)||_0 + \max_{\tau \in Y} ||f^*(\tau)||_0
\end{equation}
When a simplicial map exists, the above function has a minimum value of 2. However, it turns out that the above optimization problem is hard in a precise sense:

\begin{proposition}
Finding whether or not a bisimplicial map exists is at least as difficult as solving the graph isomorphism problem.
\begin{proof}
We use the standard technique of performing a polynomial-time reduction of an instance of the graph isomorphism problem to the bisimpliciality problem. In other words, suppose that we have an oracle that can tell us whether a bisimplicial map between two complexes exists in polynomial time. Then we must show that we can also determine whether there exists a graph isomorphism between two graphs $G$ and $H$ in polynomial time. Let us construct a machine that solves this problem in polynomial time. 

Suppose that we are given two graphs $G$ and $H$. Note that we can take care of any polynomial time graph invariants beforehand. For example this means that we may answer ``No'' if $G$ and $H$ have different numbers of vertices or edges. Similarly, since homology over a field may be computed in polynomial time (see \cite{Carlsson_04}), we may also answer ``No'' if $H_*(G) \neq H_*(H)$. Thus suppose that using our oracle we have constructed a bisimplicial map $f$ that induces an isomorphism on homology between $C_*(G)$ and $C_*(H)$.

We claim that $f$ is a graph isomorphism. Let us denote the vertices of $G$ and $H$ by $V(G)$ and $V(H)$, and the edges by $E(G)$ and $E(H)$. By bisimpliciality, $f$ must be an isomorphism between $V(G)$ and $V(H)$. Suppose that $u \sim v$ in $G$ (ie. the edge $[u, v]$ exists in $G$). Then by the fact that $f$ is a chain map, $\partial f([u, v]) = f(\partial[u, v]) = f(v - u) = f(v) - f(u)$. Thus $f([u, v])$ is an edge between $f(u)$ and $f(v)$ in $H$. So $f(u) \sim f(v)$. 

Conversely, suppose that there are vertices $x$, $y$ in $H$ such that $x \sim y$ and $f(u) = x$ and $f(v) = y$, but with $u \nsim v$. However, we must have an edge $[s, t] \in E(G)$ such that $f([s, t]) = [x, y]$ (if not $f$ would not be bisimplicial or would not be a homology-isomorphism). But then at least one of the following holds: $s \neq u$ or $t \neq v$. Without loss of generality, $s \neq u$. Then we have that $f(s) = f(u) = x$, contradicting the bisimpliciality of $f$. Thus we must have that $u \sim v$, and hence $f$ is a graph isomorphism. Thus, answering ``Yes'' when a bisimplicial map exists, and ``No'' otherwise solves the graph isomorphism decision problem.
\end{proof}
\end{proposition}

Unfortunately, it is not known whether the graph isomorphism problem is NP-complete or is in P \cite{Garey}. Thus all of the best-known algorithms are super-polynomial. Complexity theorists have created a class called GI which consists of those problems which are polynomial time reducible to the graph isomorphism problem. In this terminology, we have shown that computing bisimplicial maps is GI-hard in general. 

\subsection{The Combinatorial Approach is Hard: Empirical Findings}

Although the bisimpliciality problem is provably hard as previously shown, one may wonder whether it is possible to use heuristic combinatorial optimization techniques (over $\mathbb{Z}/2\mathbb{Z}$). Examples include random walks, simulated annealing, or greedy search with randomized restarts. Here, we give some empirical evidence suggesting that these approaches are unlikely to work for finding bisimplicial maps. Although it is possible that many simplicial maps exist for certain articially constructed cases, the examples below suggest that such maps are ``rare''. 

Consider the one of the simplest conceivable nontrivial cases: where both $X$ and $Y$ are squares containing the simplices $\{[0], [1], [2], [3], [0,1], [1,2], [2, 3], [0,3]\}$. Suppose that we are looking for a map $f: C_*(X) \rightarrow C_*(Y)$ such that $f$ is forward and backward simplicial. As stated before we wish to find a minimizer for equation (\ref{bisimplicial}).

\begin{figure}
\begin{center}
\includegraphics[width=10cm]{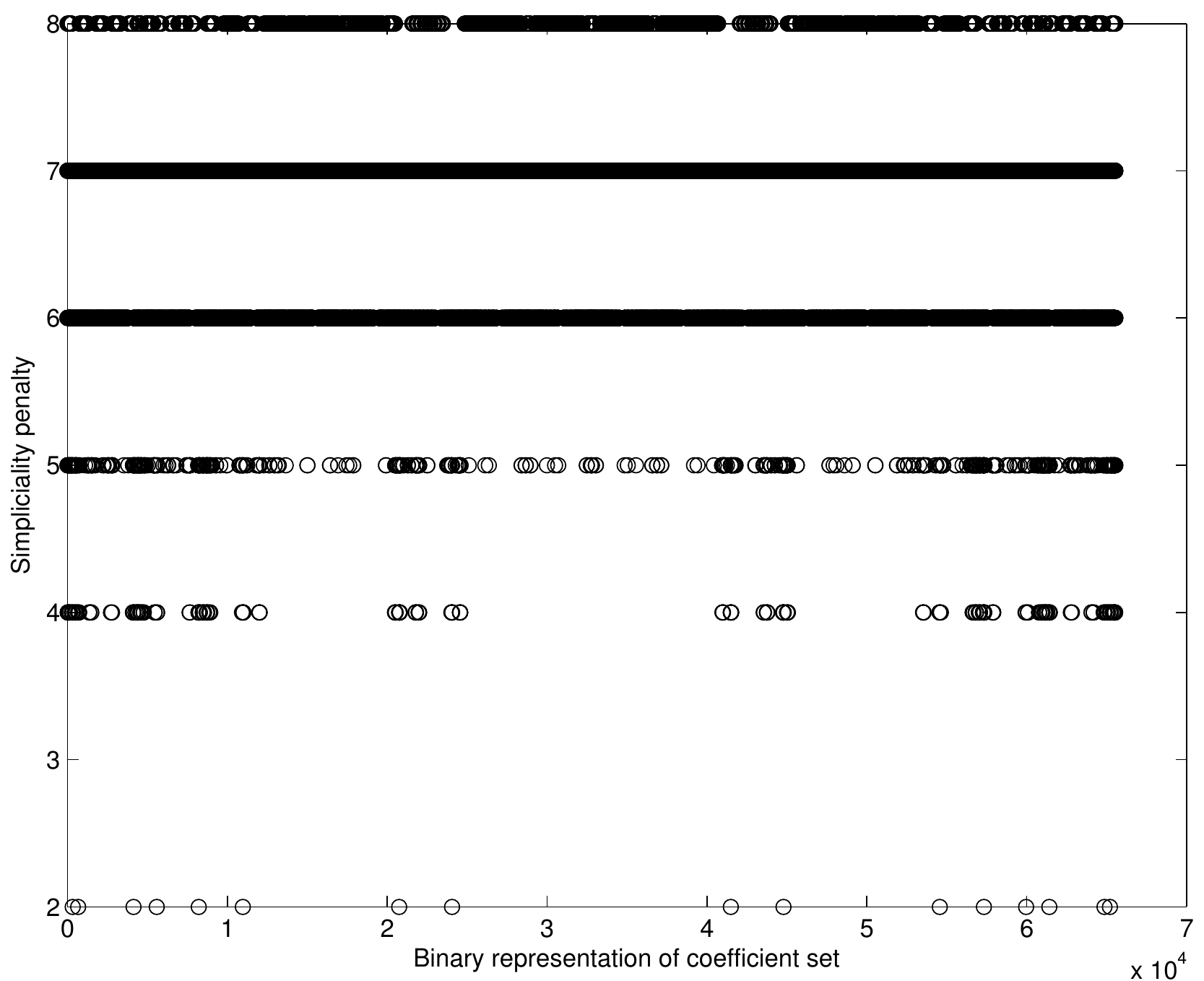}
\end{center}

\caption{Enumeration of all of the homotopy representatives of the chain map inducing an isomorphism between two squares. The horizontal axis indicates the binary representation of the set of 16 coefficients for the homotopies, and the vertical axis shows the bisimpliciality penalty.}
\label{enumeration4}
\end{figure}

The homotopies in this case consist of the images of all simple tensors of the form $[a^*] \otimes [b, c]$ under the map $d_1$. Thus there are a total of 16 homotopies. Since we are dealing with $\mathbb{Z}/2\mathbb{Z}$ coefficients, there are $2^{16} = 65,536$ possible choices for the homotopy coefficients. Although it is not practical in general, we may simply enumerate all possible sets of coefficients. If we enumerate all possible sets of coefficients, we find that out of the 65,536 possibilities, only 16 choices of coefficients yield bisimplicial maps. Figure \ref{enumeration4} shows the set of all coefficients enumerated on the horizontal axis (by using binary representation) with the simplicial objective function value on the vertical axis. 

Similarly, for finding mappings between a triangle and a square, out of the 4096 possibilities only 7 of them yield minima. Similarly, in the case where $X$ and $Y$ are circles with 8 vertices (octagons) simulated annealing was performed. After 25,300 iterations a minimum value of 11 was reported for the simplicial objective function. Since the identity map is a minimizer, the actual minimum should be 2. This was also the case with the other heuristics (greedy search and random walking) - we have yet to find a nontrivial pair in which one of these techniques is able to find a map that comes close to minimizing (\ref{bisimplicial}).  While these examples do not constitute proof, they suggest that bisimplicial maps are extremely rare among chain maps.

\subsection{Matrix Representation}

For convenience, we fix some notation that will be used for the rest of this document. Let the capital letters $F_m$ denote the matrix representations of the chain maps $f_m$, and $H_m$ denote the matrix representations of the homotopies $h_m$. We let $F = \sum_m F_m$ be the sum of chain maps. We denote an arbitrary member of $[X, Y]$ by $g$, and its matrix representation by $G$. The basis elements $\{\sigma_i\}$ of $X$ correspond to the unit vectors $\{e_i\}$, and the basis elements $\{\tau_j\}$ of $Y$ correspond to the unit vectors $\{e_j\}$.

\subsection{Minimizing Image and Preimage Size}

Let us explore the first two criteria stated at the beginning of this section. We said that ideally we would like to minimize the number of non-zero terms in the image and adjoint-image of each simplex. This can be formulated as a special case of the following optimization problem:

\begin{equation}
\label{max_optimization}
\begin{aligned}
& {\text{minimize}_{c_n}}
& & \max_{\sigma \in X} ||g(\sigma)||_p + \max_{\tau \in Y} ||g^*(\tau)||_p \\
& \text{subject to}
&& g = \sum_m b_m f_m + \sum_n c_n h_n
\end{aligned}
\end{equation}

Note that in the above expression, the coefficients $\{b_m\}$ are fixed beforehand. This means that we make an initial selection of which homological features to preserve. If they were not selected and were optimization variables, the problem above would be minimized by the zero map. 

It is also possible to replace the max terms by sums over the domain and codomain, however such a replacement yields less meaningful maps (Consider the case of the chain map which maps each vertex in the domain to a single vertex in the codomain). 

To minimize the maximum cardinality or the sum of the cardinalities, one can use $p = 0$ (although this is not actually a norm). However, we have seen that such a combinatorial approach is difficult, and since the problem dimension we are dealing with is multiplicative in the sizes of the simplicial complexes $X$ and $Y$, this is out of the question. As practiced in many optimization settings, one may relax the cardinality minimization problem to a $1$-norm minimization problem. The intuition behind this is that the unit ball in the $1$-norm is the convex hull of the points $\{\pm e_i\}$ and that constrained optima tend to lie on corner points which are sparse.

In the case of the $1$-norm, we can rewrite this as:
\begin{equation}
\label{max_optimization_matrix}
\begin{aligned}
& {\text{minimize}_{c_n}}
& & ||G||_1 + ||G^T||_1 \\
& \text{subject to}
&& G = \sum_m b_m F_m + \sum_n c_n H_n
\end{aligned}
\end{equation}

The $1$-norm of a matrix is defined to be the maximum absolute column sum of its entries, and is equal to the operator norm induced by the vector $1$-norm. It turns out that this problem is convex and in fact can be reformulated as a linear program.

Although the optimization problem (\ref{max_optimization}) with $p \geq 1$ is useful in that it eliminates many solutions which we view as being inadmissible, it still has the property that it is closed under convex combinations. This means, for example, that in the case of the two triangles, the map that sends each vertex to $\frac{1}{3} ([0] + [1] + [2])$ is an optima, but is not what we are looking for. Thus, our viewpoint changes from (\ref{max_optimization}) being the answer our search for a suitable optimization problem, to instead being a definition of admissibility of a map:

\begin{definition}
A map $g$ is said to be admissible if it is a member of the set of optima of (\ref{max_optimization}) with $p = 1$ or equivalently (\ref{max_optimization_matrix}). Denote the admissible set by $\mathcal{C}$. 
\end{definition}

The next step is to somehow identify the points within $\mathcal{C}$ that satisfy some sort of sparsity requirement. One possibility is to require the coefficients $\{c_n\}$ to be integral. This brings us into the realm of integer linear programming, which turns out to be NP-hard (in the absence of the property of integral vertices which is not satsified in this situation). In fact, the integer feasibility problem (finding a point with integer coordinates in a polytope defined by $\{x | A x \leq b\}$) is also NP-hard. Computationally, this can be applied to situations where both $X$ and $Y$ are small complexes, but it does not scale well. 

The second possibility is to optimize some measure of sparsity or peakiness over the points in $\mathcal{C}$. One strategy is to maximize the ratio of the 2-norm to the 1-norm. (To see that this is reasonable, one can consider the vectors $(1, 0 ... 0)$ and $(1/n, ... 1/n)$). Another possibility is to minimize the function which measures the distance of a point to it's nearest integral point (let the objective be $(x_1 - [x_1], ..., x_n - [x_n]$). Although both these objectives have the property that their global minima over $\mathcal{C}$ are the simlicial maps, both of them suffer from being non-convex. In fact, since the vertices of $\mathcal{C}$ are local minima of these functions, global minimization would involve searching all of the corner points. Once again, the task of enumerating vertices of a convex polytope turns out to be NP-hard \cite{Khachiyan}. A third random heuristic is to select a random search direction $v$ and solve the problem

\begin{equation}
\begin{aligned}
& {\text{minimize}}
& & v^T c \\
& \text{subject to}
&& c \in \arg \min \left( \max_{\sigma \in X} ||g(\sigma)||_p + \max_{\tau \in Y} ||g^*(\tau)||_p \right) \\
&&& g \in [X, Y]
\end{aligned}
\end{equation}
This will result in the selection of a random corner point of $\mathcal{C}$. This can be repeated until a sufficiently sparse (close to simplicial) map is found.

\subsection{The Alexander-Whitney Map}

It is easy to see that there is no convex objective function that will select all of the simplicial maps, since a convex function cannot have a disconnected set of minima. However, if we discard the requirement of unbiasedness (not favoring one simplicial map over another) or convexity we can devise other methods for selecting favorable maps.  The following method dispenses with convexity.

Define the Alexander-Whitney map $\Delta: C_*(X) \rightarrow C_*(X) \otimes C_*(X)$ by
$$
\Delta(\sigma) = \sum_{i=0}^{\deg \sigma}  \sigma|_{0 ... i} \otimes \sigma|_{i ... \deg \sigma}
$$
Where $\sigma|_{0 ... i}$ is $[v_0, ... v_i]$ if $\sigma = [v_0, ..., v_n]$, and $\sigma|_{i ... \deg \sigma}$ is $[v_i, ..., v_n]$. It is a routine calculation to show that the following diagram commutes if and only if $f$ is a simplicial chain map:
\begin{equation}
\begin{CD}
C_*(X) @>f>> C_*(Y)\\
@VV{\Delta}V @VV{\Delta}V\\
C_*(X) \otimes C_*(X) @>f \otimes f>> C_*(Y) \otimes C_*(Y)
\end{CD}
\end{equation}

Thus we can measure the deviation of a chain map from simpliciality by measuring the norm of the difference $g \otimes g(\Delta(\sigma)) - \Delta(g(\sigma))$. Suppose we are given a loss function $L$ on $\mathbb{R}^{J} \otimes \mathbb{R}^{J}$, we can define
\begin{equation}
L_{AW}(g) = \sum_{\sigma \in X} L\left(g \otimes g(\Delta(\sigma)) - \Delta(g(\sigma))\right)
\end{equation}
For example, a convenient choice would be the quadratic loss
\begin{equation}
||g||_{2, AW}^2 = \sum_{\sigma \in X} ||g \otimes g(\Delta(\sigma)) - \Delta(g(\sigma))||_2^2
\end{equation}
Thus given a selection of a loss function, we can solve
\begin{equation}
\begin{aligned}
& {\text{minimize}}
& & L_{AW}(g) + L_{AW}(g^*) \\
& \text{subject to}
&& g \in [X, Y]
\end{aligned}
\end{equation}

Note that even if the original loss function $L$ is convex, $L_{AW}$ will not be convex on the affine space of chain maps. However, it is clear the minima of the above optimization problem will be precisely the bisimplicial maps, in the case where one exists.

\begin{remark}
A naive implementation of the above would be very expensive to compute. Suppose that $|X| = I$ and $|Y| = J$, so the matrix representation of a map $f$ is a $J \times I$ matrix $F$. At some point, we need to compute a product of the form $(F \otimes F) v$. The simple way would be to form the matrix $F \otimes F$ (of size $J^2 \times I^2$), and then perform the matrix-vector multiplication. The cost of this operation is $O(I^2 J^2)$.

However, it is possible to rearrange this product differently. Suppose that $v$ is a vector in $\mathbb{F}^{I^2}$. We reshape this into the matrix $V$ of size $I \times I$ by dividing $v$ into blocks of length $I$ and laying them side by side. It turns out that computing the product $(F \otimes F) v$ is equivalent to computing $F V F^T$ and then reshaping it into vector form (see \cite{Henderson}). The complexity of this operation is $O(I^2 J)$.
\end{remark}

\section{Extensions and Applications}
\label{section_applications}

\subsection{Coordinatization on a Manifold}

In this section we investigate how the method previously described can be used to find manifold-valued coordinates for a given simplicial complex. Suppose that we have the following data:

\begin{itemize}
\item $X$: The domain simplicial complex. This is the ``original'' data set under investigation.
\item $M$: A manifold that we wish to compute coordinates on.
\item $Y$: A triangulation of the manifold $M$.
\item $\varphi$: A homeomorphism, $Y \rightarrow M$ corresponding to the triangulation of $M$.
\item $\psi$: A ``localization'' function, $C_0(Y) \rightarrow M$ which maps zero-dimensional chains on $Y$ to points on the manifold.
\end{itemize}

We enforce a compatibility constraint on functions $\varphi$ and $\psi$: If $\alpha = \sum_i c_i \sigma_i$, where $c_i = \delta_{ij}$, then $\psi(\alpha) = \varphi(\sigma_j)$. Our objective is to find a coordinate mapping $\rho: X \rightarrow M$.

The first approach to computing coordinates on $M$ is to compute the parameterization of $[X, Y]$ as well as to perform one of the optimization routines in the previous section to obtain a map $f: C_*(X) \rightarrow C_*(Y)$. Then, given a vertex $\sigma_0 \in X$, we define its coordinate on $M$ to be 
$$\rho(\sigma_0) = \psi(f(\sigma_0))$$

A second approach relies on the geometry of the manifold. Suppose that $M$ is equipped with a Riemannian metric $g$. We wish to find a mapping $X_0 \rightarrow M$ which minimizes the total distortion across the 1-skeleton of $X$. In other words, we wish that vertices connected by an edge in $X$ should be mapped nearby in $M$. This can be precisely formulated as the following optimization problem:

\begin{equation}
\label{manifold_optimization}
\begin{aligned}
& {\text{minimize}}
& & \sum_{[\sigma_i, \sigma_j] \in X_1} d^2(\psi(f(\sigma_i)), \psi(f(\sigma_j)))\\
& \text{subject to}
&& f \in [X, Y]
\end{aligned}
\end{equation}
The metric $d$ on $M$ has the standard definition:
$$d(x, y) = \inf \{L(\gamma) | \gamma: I \rightarrow M, \gamma \in C^1, \gamma(0) = x, \gamma(1) = y\}$$
and where $L(\gamma)$ is the length of the curve $\gamma$ defined by the Riemannian metric $g$.

\subsection{Example: Mapping to a circle}
\label{circlemanifold}

Let $M = \mathbb{R} / 2 \pi \mathbb{Z}$ be the circle, and suppose that $Y$ is an $n$-polygon homeomorphic to $M$. We define the coordinate of the $k$-th vertex to be $2 \pi k / n$, where $k = 0, ..., n - 1$. We also define $\psi$ to map a chain in $C_0(Y)$ to the weighted sum of its basis elements. So we have that
$$\psi(\sum_i c_i \sigma_i) = \sum_i c_i \varphi(\sigma_i)$$
Our distance function on $M$ becomes $d(x, y) = (y - x) \mod 2 \pi$. Given this setup, we can find $M$-valued coordinates for a data set of interest, $X$ by solving the optimization problem (\ref{manifold_optimization}). An example is shown in Figure \ref{manifoldmap}.

\subsection{Density Maximization}
\label{densitymaxsection}

Suppose that we have a set of points in Euclidean space, $Y_0 \in \mathbb{R}^n$ and a simplicial model $X$. We wish to find a mapping from the vertices in $X$ to Euclidean space such that the images of these vertices land in regions of high density. The interpretation is that this process produces a clustering of the data that is aware of the topological structure of the data set $Y_0$.

To do this, we begin with to pieces of data:
\begin{itemize}
\item A filtered simplicial complex, $Y$, constructed from the vertices $Y_0 \in \mathbb{R}^n$. For example, one may use the Vietoris-Rips or witness constructions discussed in \cite{Carlsson_09}.
\item An emperical density estimator on $\mathbb{R}^n$, which we call $\hat{f}(y)$ created from the data points $Y_0$. A common choice would be a kernel density estimate defined by
$$\hat{f}(y) = \sum_{i=1}^{n}\frac{1}{nh} K\left(\frac{y - y_i}{h}\right)$$
A reasonable choice for the kernel function $K$ would be the standard Gaussian density function.
\end{itemize}

Given a chain map $g:C_*(X) \rightarrow C_*(Y)$, if $\sigma \in C_0(X)$, then we interpret the image $g(\sigma) \in C_0(Y)$ to be the weighted average of the points in the chain. In other words, define $\psi(\sum c_j \tau_j) = \sum c_j \varphi(\tau_j)$, where $\varphi: Y_0 \rightarrow \mathbb{R}^n$ takes a vertex in $Y$ to its Euclidean coordinates. Since we wish to move points in the image of the chain map to regions of high-density, we form the following optimization problem

\begin{equation}
\label{density_optimization}
\begin{aligned}
& {\text{maximize}}
& & \sum_{\sigma \in X_0} \hat{f}(\psi(g(\sigma)))\\
& \text{subject to}
&& g \in [X, Y]
\end{aligned}
\end{equation}

\subsection{Mapper and Contractible Data Sets}

Another extension of the idea of hom-based mappings is to shape matching or data fusion. The discussion of gene expression data in the introduction provides the motivation for this. Suppose that we have two data sets $X_0$ and $Y_0$ which for now are just sets of points. If these data sets arise from sampling a null-homotopic space, then Proposition \ref{hclassification} tells us that the homological mapping technique described in the previous section will not be very helpful. However, there is a way to remedy this.

In the paper \cite{Mapper}, the authors describe a multiscale decomposition method called mapper. The idea is that one has a filter function $f:X \rightarrow \mathbb{R}$, and we cluster the set $X$ according to preimages of overlapping intervals. Although we do not fully describe the method here, mapper allows a statistical practitioner to obtain a multiscale representation of the data at different resolutions. The reader is advised to consult \cite{Mapper} for a detailed discussion. 

Given outputs of the mapper algorithm, we are interested in constructing maps between different representations of the same dataset, either from different filter functions or from different scale parameters. This problem is relevant in biological settings in which the obtained datasets are highly dependent on the measuring procedures used. We combine our homological mapping procedure with mapper into a structure mapping method as follows:

\begin{itemize}
\item Given two data sets $X$ and $Y$, and two filter functions $f_X, f_Y: X, Y \rightarrow \mathbb{R}$, we run the mapper algorithm to obtain reduced simplicial models $T_X$ and $T_Y$. For the 1-dimensional version of mapper on a contractible data set, $T_X$ and $T_Y$ will be trees.
\item Since the filter functions $f_X$ and $f_Y$ are also defined on the $T_X$ and $T_Y$, we take the quotient of each tree by the set of vertices which are local maxima of the filtration functions. This yields two graphs $G_X$ and $G_Y$. In general, these two graphs may have cycles.
\item We run the homological mapping algorithm to obtain an optimal chain map, $g$ between $C_*(G_X)$ and $C_*(G_Y)$.
\end{itemize}

\section{Implementation and Results}

\subsection{Software}
The above ideas were implemented as described below in a new version of the JavaPlex software package \cite{javaPlex}. Further optimization and scripting was performed using Matlab. The computation of the homology of the hom-complex was performed over the field $\mathbb{Q}$ in exact arithmetic, and the optimization was performed in floating-point.

We also note that this entire mapping procedure can be performed in a persistent setting where in addition to the natural grading of the chain complex by dimension, we have a grading by filtration. For a good discussion on persistent homology, the reader is invited to look at \cite{Carlsson_09} and \cite{Carlsson_04}. The one modification we make is that we only select representatives for $[X, Y]$ which correspond to nontrivial persistence intervals. In other words, in the expression $[X, Y] = \left\{ \sum_m b_m f_m + \sum_n c_n h_n \right\}$, we set the coefficients $b_m$ equal to 1 for significant intervals and 0 for nonsignificant intervals.

\subsection{Visualization}
\label{coloring}
The visualizations in this section show various examples of mappings between simplicial complexes. The domain complexes are on the left, and the codomains are on the right. Colors of the complexes are computed as follows:
\begin{itemize}
\item The color of the domain complex is fixed. We start with map $\mu: X_0 \rightarrow [0, 1]^3$ mapping the vertices in $X$ to their RGB values. The color of a $n$-simplex for $n > 0$ is defined to be the average of the colors of its vertices. In other words $\mu([v_0, ..., v_n]) = (n + 1)^{-1} \sum_i \mu([v_i])$.
\item To compute the color of a simplex $\tau \in Y$, under the map $f: C_*(X) \rightarrow C_*(Y)$, we define $\mu_*: Y \rightarrow [0, 1]^3$ by $\mu_*(\tau) = \mu(f^*(\tau))$, where we extend $\mu$ linearly over chains in $X$, and where $f^*$ is the adjoint of $f$. This is analogous to the definition of the pushforward of a measure.
\end{itemize}

\subsection{Examples}

\begin{figure}[p]
\begin{center}
\includegraphics[width=16cm]{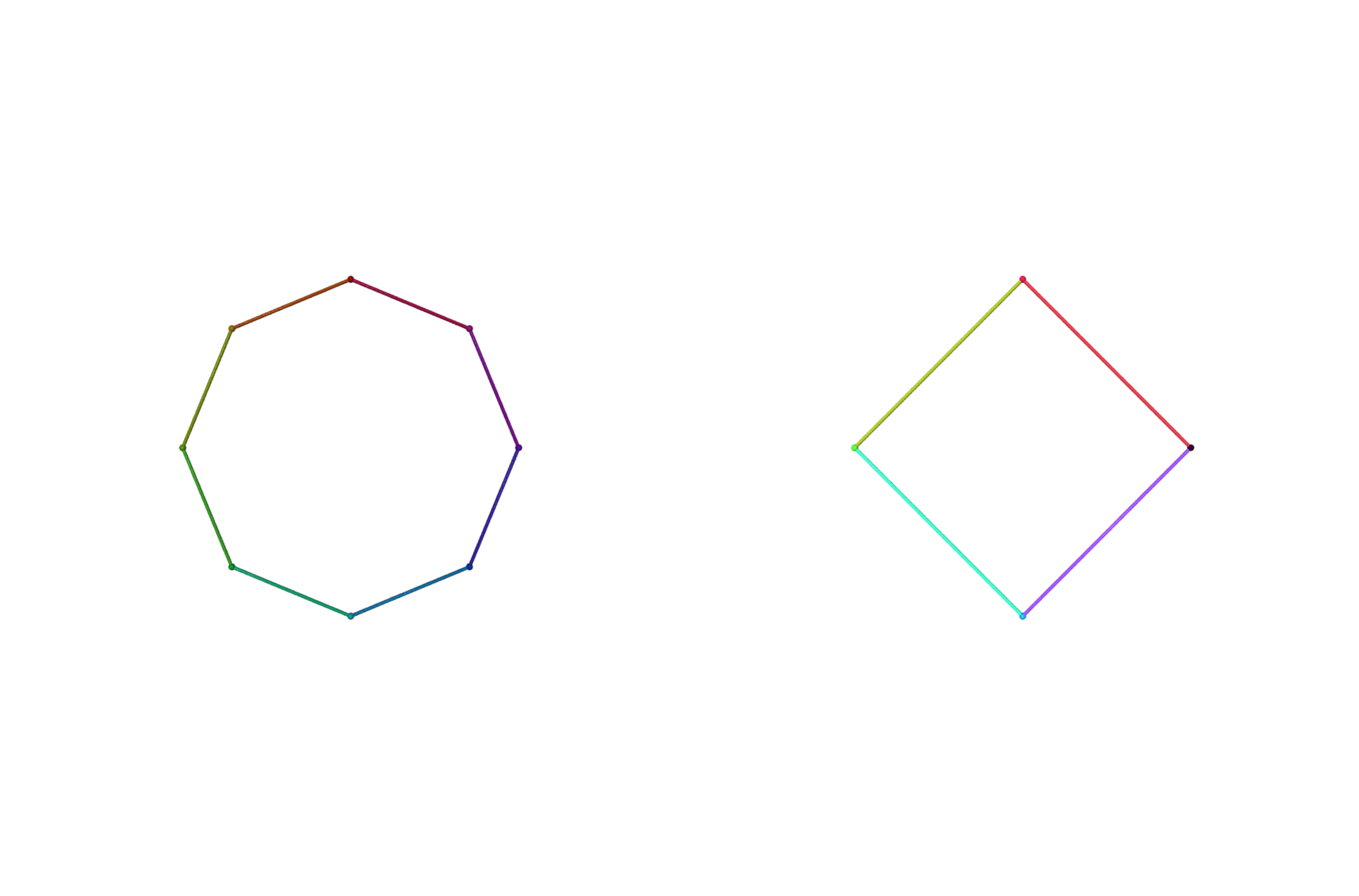}
\end{center}

\caption{Simple example of a visualization of a chain map. The map was computed by selecting a random extremal point of the polytope $\mathcal{C}$. An artefact of the visualization method described in section \ref{coloring} is that the colors on the figure on the right are more intense than those on the left. This is due to the fact that the rows in the matrix in equation \ref{map84} sum to greater than 1.}
\label{circle84}
\end{figure}

In Figure \ref{circle84}, we show an example of a homotopy representative between a circle with 8 vertices and a circle with 4 vertices. In order to compute the map, a random corner point of the admissible set, $\mathcal{C}$, was selected. In other words, the map was a random extremal point of the polytope of maps minimizing the maximum row and column sums. The computed map is given below, where the block in the upper left corner is the map on the 0-skeleton and the block in the lower right corner is the map on the 1-skeleton:
\begin{equation}
\label{map84}
\left( \begin{array}{llllllllllllllll}
1&0&0&0&0&0&0&0.66667&0&0&0&0&0&0&0&0\\
0&0&1&1&0&0&0&0&0&0&0&0&0&0&0&0\\
0&0&0&0&1&1&0&0&0&0&0&0&0&0&0&0\\
0&0&0&0&0&0&0&0.33333&0&0&0&0&0&0&0&0\\
0&0&0&0&0&0&0&0&0.33333&0&1&0.33333&0&0&0&0\\
0&0&0&0&0&0&0&0&0.66667&1&0&0&0&0&0&0\\
0&0&0&0&0&0&0&0&0&0&0&0.66667&1&0.66667&0&0\\
0&0&0&0&0&0&0&0&0&0&0&0&0&0.33333&1&1\\
\end{array} \right)
\end{equation}

\begin{figure}[p]
\begin{center}
\includegraphics[width=16cm]{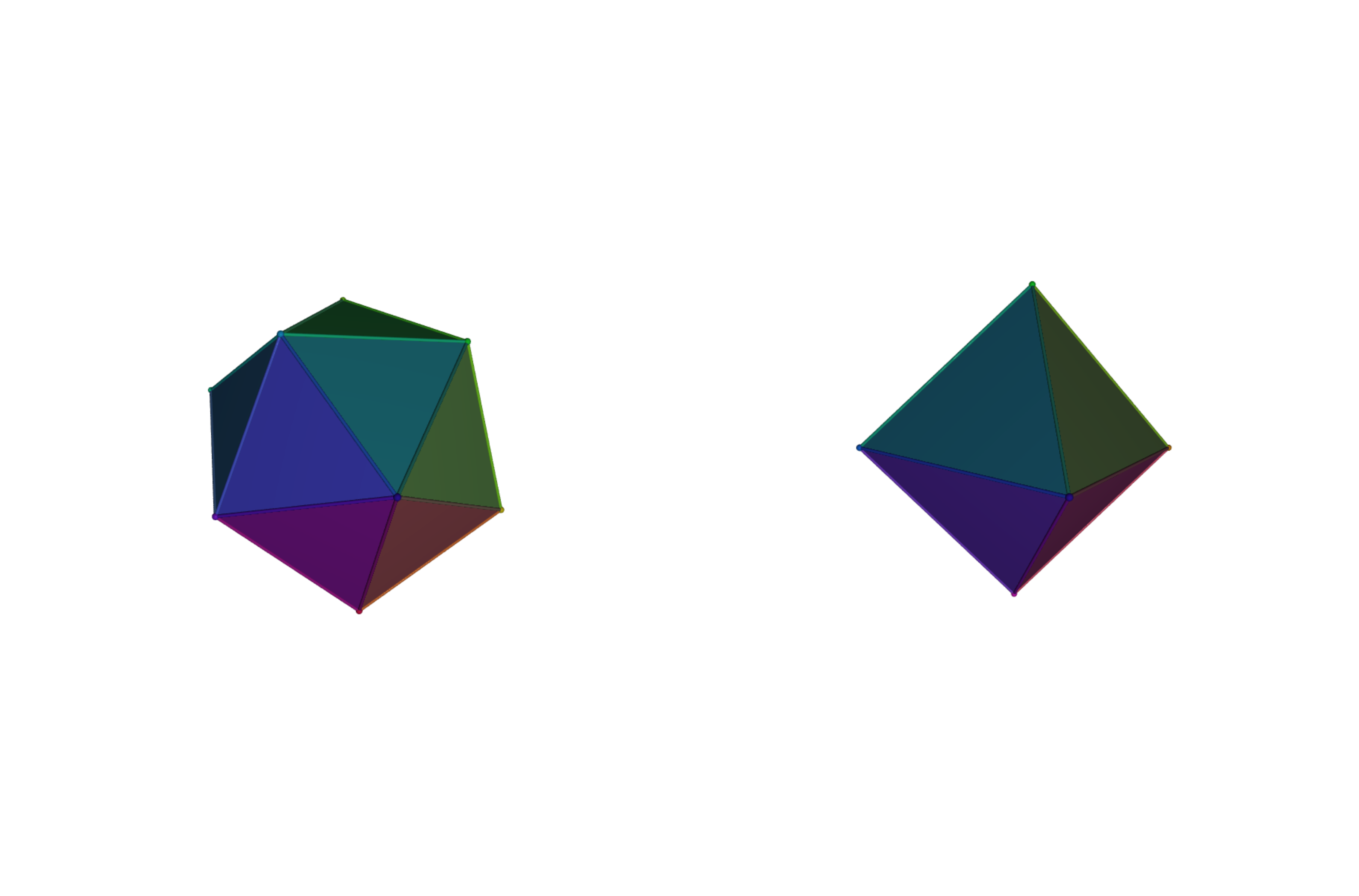}
\end{center}
\caption{This example shows an icosahedron being mapped to an octahedron. This map was constructed by performing the Alexander-Whitney optimization with quadratic loss. Note that the map was rescaled to prevent the colors in the codomain from being washed-out as in Figure \ref{circle84}.}
\label{icosafigure}
\end{figure}

\begin{figure}[p]
\begin{center}
\includegraphics[width=16cm]{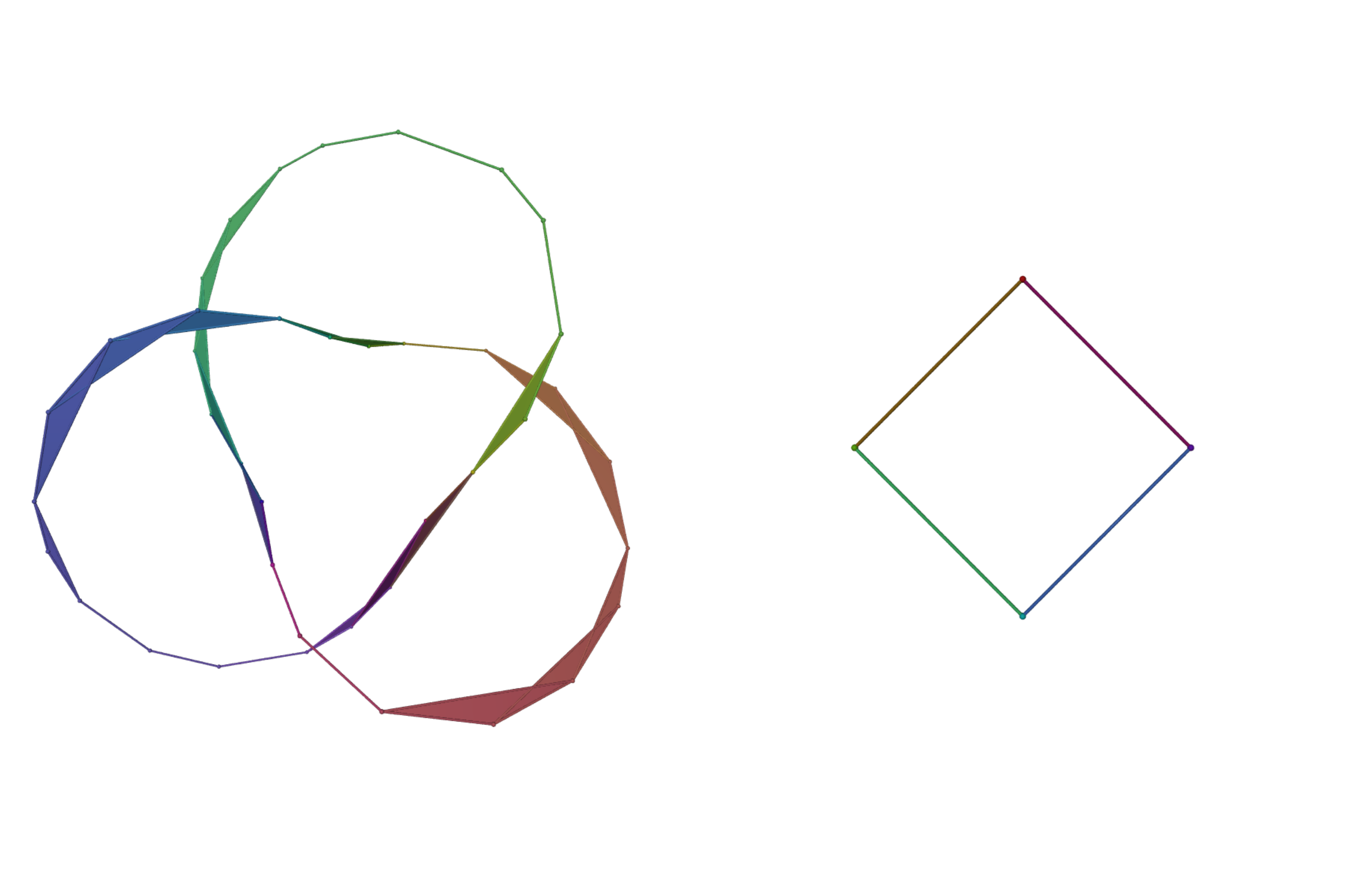}
\end{center}
\caption{The shape on the left was created by first randomly sampling 500 points on a trefoil knot. From this, the lazy-witness construction was used to construct a filtered simplicial complex as described in \cite{Witness} on a landmark set of 40 points constructed by sequential max-min selection. The mapping was obtained by randomly selecting 100 vertices of the polytope $\mathcal{C}$, and then choosing the one which had greatest 2-norm to 1-norm ratio.}
\label{trefoilfigure}
\end{figure}

\begin{figure}[p]
\begin{center}
\includegraphics[scale=0.7]{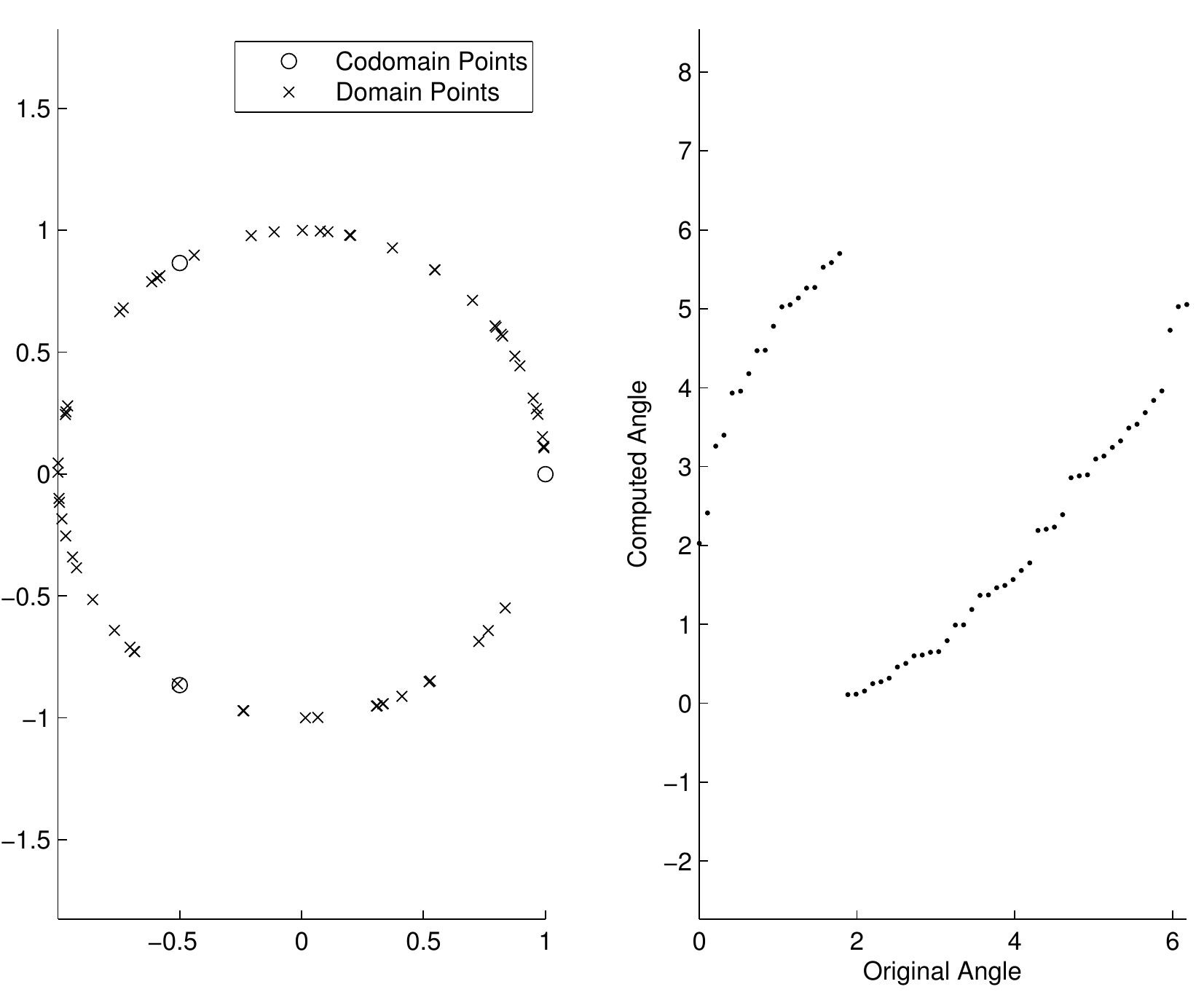}
\end{center}
\caption{Manifold map example. The domain, $X$, consists of a simplicial circle with 60 vertices, and the codomain, $Y$, consists of a circle embedded in the plane. Given a map $f: X \rightarrow Y$, we may compute the embedded coordinates for the points in $X$ as described in section \ref{circlemanifold}. The object here is to find a map from $X$ to $Y$ that minimizes the total distortion across the 1-skeleton of the domain. On the left, the images of the domain points are shown as crosses, whereas the codomain points are shown as circles. On the right, we show the relationship between the original angular coordinates for points in the domain on the horizontal axis, versus the computed angular coordinates on the vertical axis.}
\label{manifoldmap}
\end{figure}

\begin{figure}[p]
\begin{center}
\includegraphics[scale=0.35]{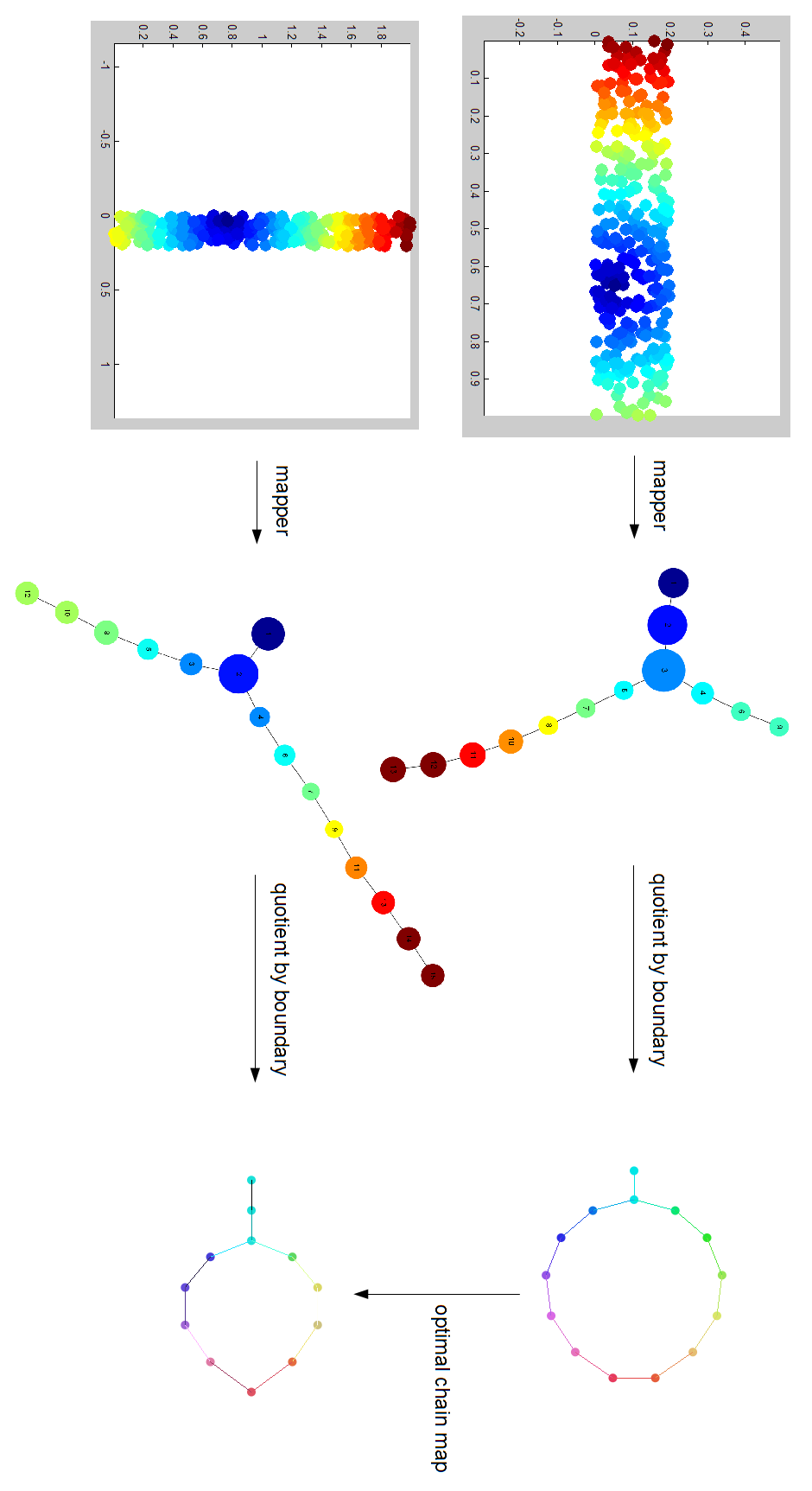}
\end{center}
\caption{This figure shows the homological mapping algorithm applied to mapper outputs. The mapper outputs are quotiented out by maxima of the filtration function. For this example, we used an eccentricity filter.}
\label{mapperfigure}
\end{figure}

In Figure \ref{icosafigure} we show an example of a map computed by minimizing the Alexander-Whitney function with quadratic loss. In Figure \ref{trefoilfigure}, the figure on the left is a simplicial complex created using the lazy-witness construction from a sample of 500 points on a trefoil knot, \cite{Witness}. A map to a circle with 4 vertices is shown. In Figures \ref{manifoldmap} - \ref{mapperfigure}, we explore the applications described in section \ref{section_applications}: computing a map to a manifold, density maximization, and contractible datasets. We refer the reader to the captions for more information regarding these.

\begin{figure}[p]
\begin{center}
\includegraphics[scale=0.7]{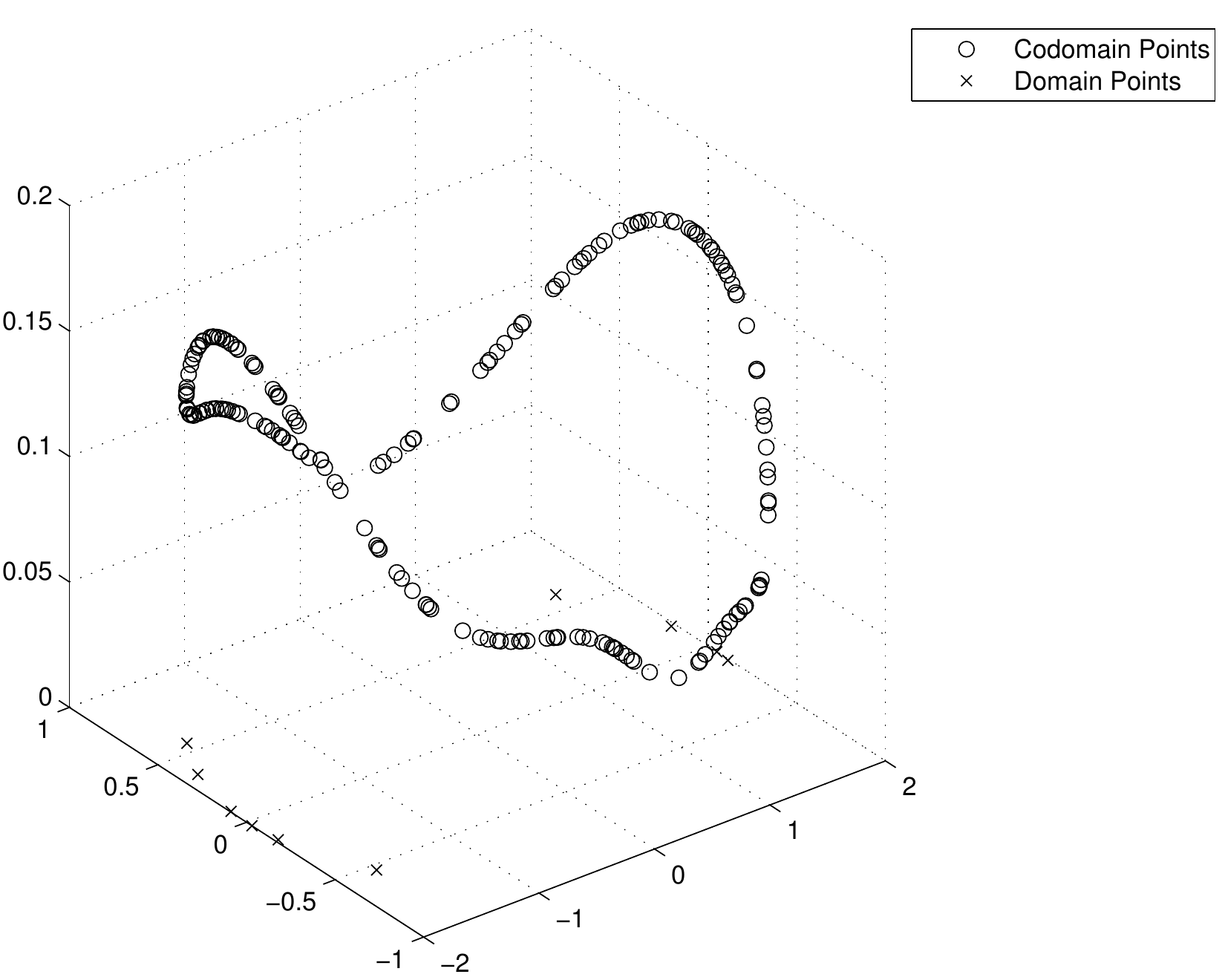}
\end{center}
\caption{Density maximization example. In the above figure, the codomain points consist of a random sample from the unit circle, and the domain complex is an idealized circle with 10 vertices. The locations of the domain points are computed by selecting the homotopy representative that maximizes the density of the image points as described in section \ref{densitymaxsection}.}
\end{figure}

\section{Concluding Remarks}

In this paper we have discussed a method for computing maps between two simplicial complexes that respect their homological structure. The computation is done in a two-stage process: first a parameterization is obtained for the homotopy classes of chain maps, and then an optimization procedure is run to select one of the maps from the affine parameterization. We have also demonstrated the method on various examples. Some key distinguishing features in comparison with traditional statistical dimensionality reduction and mapping techniques include:

\begin{itemize}
\item The domain and codomain data sets are not required to be Euclidean spaces, or even metric spaces. 
\item Conventional linear and nonlinear dimensionality reduction methods rely on the fact that the data can be somehow unfolded into a convex subset of Euclidean space. The homological method presented is designed to preserve nontrivial topological structure.
\item Unlike the method of circular coordinates, or various other surface mapping algorithms, in principle the method presented in this paper is not restricted by the dimension or structure of either the domain or codomain spaces.
\end{itemize}

Nevertheless, the mapping technique in its current state suffers from a few shortcomings:

\begin{itemize}
\item There is no universal optimization problem which produces geometrically satisfying maps in all cases. Depending on the application or situation, a practitioner might want to use different objective functions or constraints.
\item Since the computation relies on the construction of the hom-complex, the fundamental problem size given simplicial complexes of sizes $I$ and $J$ is the product, $IJ$. This leads to somewhat poor algorithmic complexity in comparison with first order methods and has limited the sizes of the examples presented.
\end{itemize}

A key step to improving the applicability of hom-complex based mappings would be to alleviate the problems with its algorithmic efficiency. It would be interesting to investigate what optimizations would enable this method to scale to datasets of more reasonable size. Despite these shortcomings, the examples in this paper are designed to be a proof-of-concept for hom-complex based mappings. 

\bibliographystyle{amsalpha}
\bibliography{biblio}

\end{document}